\documentclass[aps,pra,reprint,showpacs,superscriptaddress]{revtex4-1}

\usepackage[utf8]{inputenc}
\usepackage{amsmath}
\usepackage{bm}
\usepackage{graphicx}
\usepackage{siunitx}
\usepackage{hyperref}
\newcommand{\braket}[1]{\ensuremath{\langle #1 \rangle}}

\renewcommand{\sin}[1]{\textrm{sin}\left[#1\right]}

\begin{document}
\title{Local correlations reveal the superfluid to normal boundary in a trapped two-dimensional quantum gas}
\author{Romain Dubessy}
\author{Camilla De Rossi}
\author{Mathieu de Go\"er de Herve}
\affiliation{Universit\'e Paris 13, Sorbonne Paris Cit\'e, Laboratoire de physique des lasers, F-93430, Villetaneuse, France}
\affiliation{CNRS, UMR 7538, F-93430, Villetaneuse, France}
\author{Thomas Badr}
\author{Aur\'elien Perrin}
\affiliation{CNRS, UMR 7538, F-93430, Villetaneuse, France}
\affiliation{Universit\'e Paris 13, Sorbonne Paris Cit\'e, Laboratoire de physique des lasers, F-93430, Villetaneuse, France}
\author{Laurent Longchambon}
\affiliation{Universit\'e Paris 13, Sorbonne Paris Cit\'e, Laboratoire de physique des lasers, F-93430, Villetaneuse, France}
\affiliation{CNRS, UMR 7538, F-93430, Villetaneuse, France}
\author{H\'el\`ene Perrin}
\affiliation{CNRS, UMR 7538, F-93430, Villetaneuse, France}
\affiliation{Universit\'e Paris 13, Sorbonne Paris Cit\'e, Laboratoire de physique des lasers, F-93430, Villetaneuse, France}

\begin{abstract}
This paper reports the \emph{model free} determination of the two-fluid dynamics in a trapped two-dimensional Bose gas, relying on a local principal component analysis of the dynamics after a sudden excitation.
\end{abstract}
\maketitle

\section{Introduction}
Trapped ultracold atomic quantum gases are ideal tools to study complex quantum phenomena. By using very anisotropic traps it is possible to reduce the dimensionality of such a system and therefore perfoms experiments with degenerate Bose or Fermi gases in one or two dimensions. This has lead to the study of the Berezinskii-Kosterlitz-Thouless (BKT) transition in inhomogeneous ensembles, and in particular the measurement of the gas equilibrium properties~\cite{Rath2010,Yefsah2011} and equation of state~\cite{Hung2011}. The subtle interplay between the BKT physics and Bose-Einstein condensation has also been studied by varying the interaction strength between bosonic atoms~\cite{Fletcher2015}.
The superfluid properties of these systems have been evidenced using both a local measurement of the gas critical velocity~\cite{Desbuquois2012} and a global collective oscillation~\cite{DeRossi2016}. 

Recently we have shown that the study of the scissors mode in a finite temperature trapped Bose gas requires a local average analysis to reveal the dynamics of the normal and superfluid part of the gas~\cite{DeRossi2016}. This approach relies on a specific model predicting the expected damping rates and oscillation frequencies of an interacting classical gas, allowing to fit the data with a single parameter, the collision rate $\tau$. This model works also surprisingly well to describe the scissors oscillations of a superfluid, both in terms of frequencies and damping rates, although it relies on the assumption that the superfluid dynamics can be adequately modeled by the dynamics of a classical gas in the hydrodynamic regime.
In contrast we have also developed \emph{model free} methods to identify the collective excitations of degenerate Bose gases, based on principal component analysis (PCA)~\cite{Dubessy2014}. This method allows to extract the low energy collective excitations of the gas from the analysis of its out of equilibrium density-density correlations. For example it gives access to the spatial shape of a specific excited mode as well as its oscillation frequency, even when several collective modes are exited~\cite{Dubessy2014}. In the low temperature limit, when the thermal fraction vanishes, it can be shown that it allows to recover the exact Bogoliubov modes of the gas~\cite{Dubessy2014}. Moreover, PCA provides filtering of the data from technical imaging noise~\cite{Segal2010,Chiow2011,Farkas2014}.

In this paper we show how PCA can be extended to a local analysis, resulting in a \emph{model free} determination of the coexistence of two phases in a trapped finite temperature two-dimensional Bose gas. These two phases can be attributed to the superfluid part and normal part of the inhomogeneous Bose gas. This work is particularly relevant for the study of finite temperature Bose gas out-of equilibrium, for which it is difficult to get a self consistent and accurate description of the dynamics.

\section{Results}
We prepare a degenerate Bose gas in a radio-frequency dressed quadrupole trap~\cite{Merloti2013a}, loaded from an hybrid trap~\cite{Dubessy2012a}. By tuning the magnetic gradient and the radio-frequency amplitude and polarization we first prepare a sample confined in an oblate harmonic trap with frequencies $(\omega_x,\omega_y,\omega_z)=2\pi\times(34,48,1830)$ Hz, with a small in-plane anisotropy $\epsilon=(\omega_y^2-\omega_x^2)/(\omega_y^2+\omega_x^2)=0.34$. We then suddenly rotate the horizontal anisotropy by $\sim10^\circ$, as described in reference~\cite{DeRossi2016}. We analyze the gas dynamics by recording in-situ absorption images of the out of equilibrium density profile. Here we focus on a sample where the local average analysis~\cite{DeRossi2016} evidences a different dynamical response for the central, denser, part of the cloud and for the outer part. This can be intuitively understood in the context of the local density approximation: due to the harmonic confinement, the central part of the gas is dense enough to be superfluid, while the outer part is not, hence the different dynamical behavior.

\begin{figure}[t]
\includegraphics[width=\linewidth]{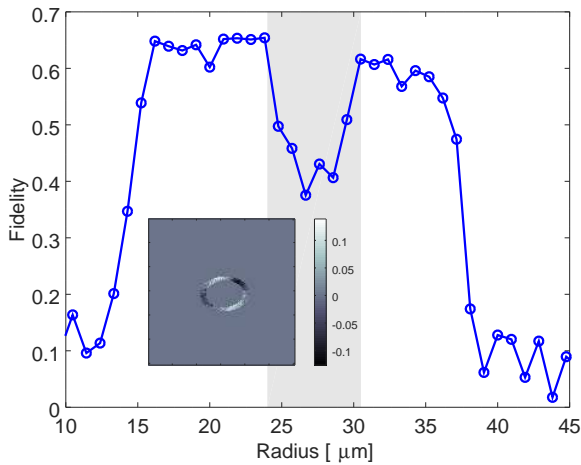}
\caption{\label{fig1}Fidelity (as defined in equation \ref{eqn:fidelity}) as a function of the region of interest radius. The inset shows the typical principal component profile $n_{\rm PC}^{(1)}$ (here for a region of interest with a radius of 15~$\mu$m). The field of view is 115~$\mu$m. See text for details}
\end{figure}

Extending now our method of local average analysis \cite{DeRossi2016}, we study the density-density correlations over a region of interest (ROI) corresponding to a constant equilibrium density $n_{\rm eq}(\bm{r})$. As the local equilibrium density depends on $\bm{r}$ only through the local chemical potential, $\mu_{\rm loc}(\bm{r})=\mu_0-V_{\rm trap}(\bm{r})$, this ROI takes the form of an ellipsoid, whose major and minor axes are aligned respectively with the weak and strong in-plane trap axis, $x$ and $y$. By applying the PCA on this ROI we extract the density profiles $n_{\rm PC}$ of the principal components, see the inset of Fig.~\ref{fig1}. We then sort the components using their overlap with the expected density pattern $\delta n_{sc}$ for the scissors oscillations: $\left|\braket{n_{\rm PC}|\delta n_{sc}}\right|$~\footnote{As the observable for the scissors mode is the $\braket{xy}$ average, both in the superfluid and normal phases, we compute the overlap with a $\sin{2\phi}$ density profile, where $\phi$ is the angle of the cylindrical coordinates.}.
In order to quantify whether a single, well defined, component matches the spatial structure of the scissors mode we define the fidelity as:
\begin{equation}
\mathcal{F}=\sqrt{\left|\braket{n_{\rm PC}^{(1)}|\delta n_{sc}}\left(\braket{n_{\rm PC}^{(1)}|\delta n_{sc}}-\braket{n_{\rm PC}^{(2)}|\delta n_{sc}}\right)\right|},
\label{eqn:fidelity}
\end{equation}
where $n_{\rm PC}^{(1)}$ and $n_{\rm PC}^{(2)}$ are the first two components closest to the scissors mode spatial shape.
The right hand side of equation~\ref{eqn:fidelity} is maximized when a single component is close to the target profile and vanishes when the overlap is small or when there are two close candidates.

Figure~\ref{fig1} displays the fidelity as a function of the radius of analysis, for a 4~$\mu$m wide ellipsoidal ROI\footnote{We have checked that the choice of the ROI width in the range $2-6~\mu$m has little influence on the results. The value $4~\mu$m is consistent with our optical resolution.}.
We first note that the PCA evidences a well defined mode for each radius in the range $16~\mu{\rm m}\leq r\leq 37~\mu{\rm m}$, therefore confirming that a scissors collective oscillation is indeed excited in the cloud.
As expected for a surface mode~\cite{Anglin2001}, we find that for too small radii there is no clear signature of the scissors oscillation. For the largest radii the fidelity drops because the signal to noise ratio in the pictures decreases as the local density is reduced.
Interestingly, the fidelity plotted in Fig.~\ref{fig1} presents a dip at intermediate radii (gray shaded area), suggesting that the scissors-like correlations are locally reduced.

To investigate this more precisely, we turn now to the study of the correlations between different radii. For a given radius $r_n$, we select the principal component with the highest fidelity PC$_n$ (as displayed on Fig.~\ref{fig1}), and compute its time dependent weight $c_{n}(t)$ for the data set. We note that when the PCA is used on the whole sample, this weight gives access to the oscillation frequency of the associated collective mode~\cite{Dubessy2014}. We then compute the correlation matrix between the weights at different radii:
\begin{equation}
\mathcal{C}(r_n,r_k)=\sum_{i}c_{n}(t_i)c_{k}(t_i).
\label{eqn:correlation}
\end{equation} 

\begin{figure}[t]
\includegraphics[width=\linewidth]{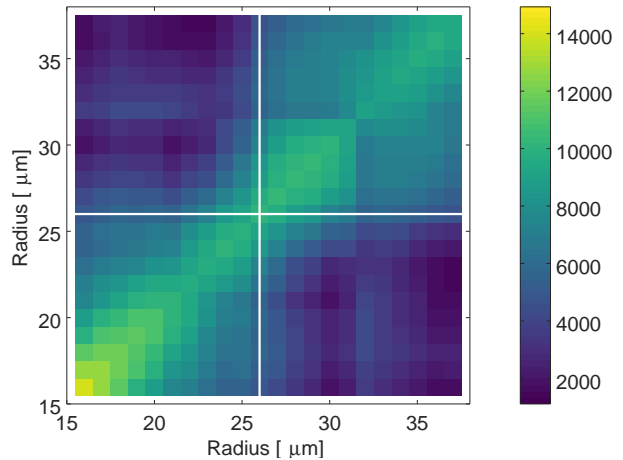}
\caption{\label{fig2}Visualization of the correlation matrix defined in equation~\ref{eqn:correlation}. The color scale indicates the magnitude of the absolute value of the matrix elements, using arbitrary units. The thin white lines are guides to the eye to identify the areas where the correlations are significant.}
\end{figure}
Figure~\ref{fig2} displays a color plot of the magnitude of the correlation matrix elements. As underlined by the white lines, the structure of the correlations reveals that the inner part of the gas is weakly correlated to the outer part. As we study out-of equilibrium time correlations, this leads to the conclusion that the dynamics of the inner and outer regions of the gas are independent, and that a two-fluid model must be used to describe the dynamics. Note that we plotted the matrix element values only for the range of radii where the fidelity is significant, see Fig.~\ref{fig1}.

\section{Discussion}
This work shows that the PCA and its extension to a local analysis is a very powerful tool to extract information, without a priori knowledge of the dynamics. 
By focusing on principal components with a given spatial pattern, corresponding to the expected scissors mode profile, we find that a well defined mode can be identified for a range of radii, using our modified definition of the fidelity. By studying the correlations between the principal components at different radii, we show that the scissors oscillation in the inner region of the fluid is uncorrelated to the oscillation in the outer region. This proves that a two-fluid dynamics is at play in the gas. Here we identify the superfluid component by comparing the measured oscillations frequencies to a model~\cite{DeRossi2016}, or the local equilibrium density to the local density approximation prediction~\cite{Prokofiev2001}. For our data we find that the weights computed at radii lower than 26~$\mu$m oscillate with a frequency close to that of a superfluid $\omega_{\rm sc}=\sqrt{\omega_x^2+\omega_y^2}$, hence revealing the
superfluid boundary.

We note that the boundary between the two fluids found with the local PCA occurs at a radius $4~\mu$m larger than the one where the BKT boundary is measured by the local average analysis~\cite{DeRossi2016}. The reason for this discrepancy is not yet fully understood.
It will be interesting to study with this local PCA approach other collective modes of the gas~\cite{Dubessy2014}, in the presence of a non negligible thermal fraction.

\begin{acknowledgments}
We acknowledge financial support from ANR project SuperRing (ANR-15-CE30-0012-01).
LPL is a member of DIM SIRTEQ (Science et ing\'enierie en R\'egion \^Ile-de-France pour les technologies quantiques).
\end{acknowledgments}

\end{document}